\documentclass{emulateapj}

\usepackage{natbib}
\usepackage{amsmath}
\usepackage{url}
\usepackage{longtable}

\newcommand{\bvri}{\protect\hbox{$BV\!RI$} }

\newcommand{\about}{$\sim\!\!$~}
\newcommand{\kms}{\,km\,s$^{-1}$}

\newcommand{\err}[2]{\ensuremath{^{+#1}_{-#2}}}

\def\lsim{\hbox{\rlap{\raise 0.425ex\hbox{$<$}}\lower 0.65ex\hbox{$\sim$}}}
\def\gsim{\hbox{\rlap{\raise 0.425ex\hbox{$>$}}\lower 0.65ex\hbox{$\sim$}}}

\def\arcsec{\hbox{$^{\prime\prime}$}}
\def\arcdeg{\mbox{$^\circ$}}

\shorttitle{Metallicity Differences in SN~Ia UV Spectra}
\shortauthors{Foley \& Kirshner}

\begin{document}

\title{Metallicity Differences in Type I\lowercase{a} Supernova
  Progenitors Inferred from Ultraviolet Spectra}

\def\cfa{1}

\author{
{Ryan~J.~Foley}\altaffilmark{\cfa}, and
{Robert~P.~Kirshner}\altaffilmark{\cfa}
}

\altaffiltext{\cfa}{
Harvard-Smithsonian Center for Astrophysics,
60 Garden Street, 
Cambridge, MA 02138, USA
}

\begin{abstract}
  Two ``twin'' Type Ia supernovae (SNe~Ia), SNe~2011by and 2011fe,
  have extremely similar optical light-curve shapes, colors, and
  spectra, yet have different ultraviolet (UV) continua as measured in
  \textit{Hubble Space Telescope} spectra and measurably different
  peak luminosities.  We attribute the difference in the UV continua
  to significantly different progenitor metallicities.  This is the
  first robust detection of different metallicities for SN~Ia
  progenitors.  Theoretical reasoning suggests that differences in
  metallicity also lead to differences in luminosity.  SNe~Ia with
  higher progenitor metallicities have lower $^{56}$Ni yields, and
  lower luminosities, for the same light-curve shape.  SNe~2011by and
  2011fe have different peak luminosities ($\Delta M_{V} \approx
  0.6$~mag), which correspond to different $^{56}$Ni yields: $M_{\rm
    11fe}(^{56}{\rm Ni}) / M_{\rm 11by}(^{56}{\rm Ni}) =
  1.7\err{0.7}{0.5}$.  From theoretical models that account for
  different neutron to proton ratios in progenitors, the differences
  in $^{56}$Ni yields for SNe~2011by and 2011fe imply that their
  progenitor stars were above and below solar metallicity,
  respectively.  Although we can distinguish progenitor metallicities
  in a qualitative way from UV data, the quantitative interpretation
  in terms of abundances is limited by the present state of
  theoretical models.
\end{abstract}



\section{Introduction}\label{s:intro}

Type Ia supernovae (SNe~Ia) are extremely useful for measuring cosmic
distances because they are intrinsically luminous, and after
standardization \citep[e.g.,][]{Phillips93}, have very small scatter
in their intrinsic luminosities.  Because of these qualities, SNe~Ia
were used to discover the Universe's acceleration
\citep{Riess98:Lambda, Perlmutter99} and are useful for measuring
cosmological parameters \citep[e.g.,][]{Suzuki12}.  These measurements
assume that SNe~Ia across all redshifts, and thus cosmic time, have
the same luminosity after standardization by light-curve shape.

Recent observations suggest ways to improve this paradigm.  SNe~Ia in
late-type host galaxies have systematically larger peak luminosities
than those in early-type galaxies \citep[e.g.,][]{Hamuy96:lum}.  Those
differences mostly disappear after correcting for light-curve shapes.
But even after that correction, there is a slight difference in the
luminosities of SNe~Ia coming from high-mass, passive host galaxies
and low-mass, star-forming galaxies \citep{Kelly10, Lampeitl10:host,
  Sullivan10}.  This effect is small but real, amounting to a
difference of \about 0.08~mag compared to the \about 0.2~mag effect of
acceleration relative to a matter-only universe at $z \approx 0.5$.
Even with the lowest scatter measurements \citep[e.g.,][]{Mandel11}
and correcting for additional parameters \citep[e.g.,][]{Foley11:vel},
intrinsic scatter remains.  Present methods for correcting
luminosities do not account for all the physical diversity among
SNe~Ia.

Additionally, SNe~Ia from progenitor systems with winds or outflows
tend to have higher ejecta velocities than those that have no
indication of winds or outflows \citep{Foley12:csm}.  This
observational difference may presage variance in peak luminosity.
Other subtle differences in the observables may relate to the SN
progenitor system.  \citet{Timmes03} predicted that SNe~Ia from
progenitors that differ by metallicity alone will produce different
amounts of $^{56}$Ni, and emit different peak luminosities.  Higher
metallicity results in additional neutrons, which produce more stable
Fe-group elements and less radioactive $^{56}$Ni.  \citet{Mazzali06}
found that altering the progenitor metallicity changes the peak
luminosity of SNe~Ia, but it would not change the light-curve shape.
As a result, the light-curve shape correction would be insufficient to
perfectly calibrate SN~Ia luminosities, and would result in increased
scatter.  If progenitor metallicity changes (on average) with
redshift, SN~Ia distances will be systematically biased with redshift.

Theoretical studies indicate that varying the metallicity of a SN~Ia
progenitor does not significantly affect the optical spectral-energy
distribution (SED) of a SN~Ia, but will dramatically change the UV SED
\citep[e.g.,][]{Hoflich98, Lentz00, Sauer08, Walker12}.  Increased
progenitor metallicity changes the final composition of the SN ejecta
and increases the line blanketing in the UV.  As a result, we expect a
correlation between the UV SED of a SN~Ia and its peak luminosity.

There have also been several observational efforts to obtain UV
spectra of SNe~Ia \citep[e.g.,][]{Jeffery92, Foley08:uv, Sauer08,
  Foley12:09ig, Maguire12}.  However, most spectra from these studies
do not probe below 2500~\AA, have a poor signal-to-noise ratio (S/N),
or were obtained significantly after maximum brightness.  The
exceptions are SNe~2011by (presented here), 2011fe \citep{Foley13:ca}
and 2011iv\citep{Foley12:11iv}, which all have excellent maximum-light
UV spectra obtained by \textit{HST}.

In this Letter, we forge the observational link between the UV SED of
SNe~Ia and their peak luminosities, confirming the theoretical
expectation.  We examine \textit{Hubble Space Telescope}
(\textit{HST}) STIS spectra of two SNe~Ia: SNe~2011by and 2011fe.  We
present these data in Section~\ref{s:obs}.  Besides SN~2011iv, these
are the only maximum-light spectra of SNe~Ia with a reasonable S/N
that probe wavelengths $<$2800~\AA.  Conveniently, both SNe have
negligible host-galaxy reddening and nearly identical optical colors,
light-curve shapes, and spectra.  In Section~\ref{s:anal}, we show
that the SNe have different UV spectra which implies that the
progenitor of SN~2011by had a higher metallicity than SN~2011fe.  In
Section~\ref{s:conc}, we discuss the implications of this result.


\section{Observations and Data Reduction}\label{s:obs}

SN~2011by was discovered by \citet{Jin11} on 2011 April 26.8 (UT used
throughout) in NGC~3972, an Sbc galaxy with $D = 18.5$~Mpc ($\mu =
31.34 \pm 0.36$~mag) from a Tully-Fisher measurement \citep{Tully09}.
\citet{Zhang11} obtained an optical spectrum of SN~2011by on 2011
April 27.5, only 0.7~days after discovery, and determined that it was
a young SN~Ia.

\citet{Silverman13} published optical light curves and spectra.  The
SN is spectroscopically normal with minimal dust reddening.
\citet{Silverman13} determined that SN~2011by reached maximum
brightness in the $B$ band on 2011 May 9.9 and had $\Delta m_{15} (B)
= 1.14 \pm 0.03$~mag.  \citet{Maguire12} reported that SN~2011by
peaked on 2011 May $9.6 \pm 0.1$.  Throughout this Letter, we use the
\citet{Maguire12} value.  \citet{Johansson12} presented
\textit{Herschel} data which indicates that there was a minimal amount
of circumstellar dust ($M_{\rm dust} \lesssim 0.1$~$M_{\sun}$).

SN~2011by was observed by \textit{HST} using the STIS spectrograph
(Program GO--12298; PI Ellis) on 2011 May 9.36, corresponding to $t =
0.1$~days relative to $B$ maximum.  The spectra were obtained with two
different gratings and the $52\arcsec \times 0.\arcsec2$ slit.  Two
exposures were obtained for each of the MAMA/G230L and CCD/G430L
setups with total exposure times of 5316 and 2263~s, respectively.
The two setups yield a combined wavelength range of 1605 -- 4695~\AA.
We retrieved the data from the Mikulski Archive for Space Telescopes.
The data were reduced using the standard \textit{HST} Space Telescope
Science Data Analysis System (STSDAS) routines to bias subtract,
flat-field, extract, wavelength-calibrate, and flux-calibrate each SN
spectrum.  \citet{Maguire12} presented only the G430L spectrum.  We
combine the \textit{HST} spectrum with an optical spectrum obtained
less than one day later (on 2011 May 10.24) from \citet{Silverman13}.
This extends the spectrum to 10,196~\AA.

SN~2011fe was discovered by the Palomar Transient Factory
\citep{Nugent11:atel} in M101, an Scd at $D = 6.4$~Mpc ($\mu = 29.04
\pm 0.05$~mag; \citealt{Shappee11}).  Various studies
\citep[e.g.,][]{Li11:11fe, Nugent11, Bloom12, Horesh12, Johansson12,
  Margutti12, Patat12} all indicate that SN~2011fe had negligible
circumstellar and interstellar dust reddening and the companion star
was either a white dwarf or a low-mass non-degenerate star.  Several
groups have determined that SN~2011fe had $\Delta m_{15} (B) =
1.10$~mag \citep[e.g.,][]{Richmond12}.  An \textit{HST} spectrum of
SN~2011fe (Program GO--12298; PI Ellis) with a phase of $t = 0.0$~d
relative to $B$-band maximum brightness was presented by
\citet{Foley13:ca}.  We examine this spectrum further here.

\begin{figure*}
\begin{center}
\epsscale{1.15}
\rotatebox{0}{
\plottwo{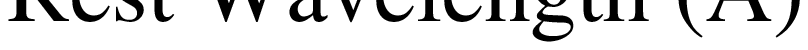}{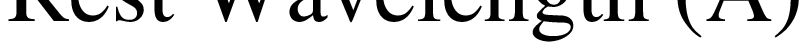}}
\caption{Maximum-light spectra of SNe~2011by (blue) and 2011fe
  (black).  The left and right panels show the same spectra, but with
  logarithmic and linear flux scales, respectively, and slightly
  different wavelength ranges.}\label{f:by_fe}
\end{center}
\end{figure*}

The spectra were dereddened by the Galactic values of
\citet{Schlafly11}: $E(B-V) = 0.012$ and 0.008~mag for SNe~2011by and
2011fe, respectively.  We display the spectra in Figure~\ref{f:by_fe}.

\section{Analysis}\label{s:anal}

\begin{figure}
\begin{center}
\epsscale{1.65}
\rotatebox{90}{
\plotone{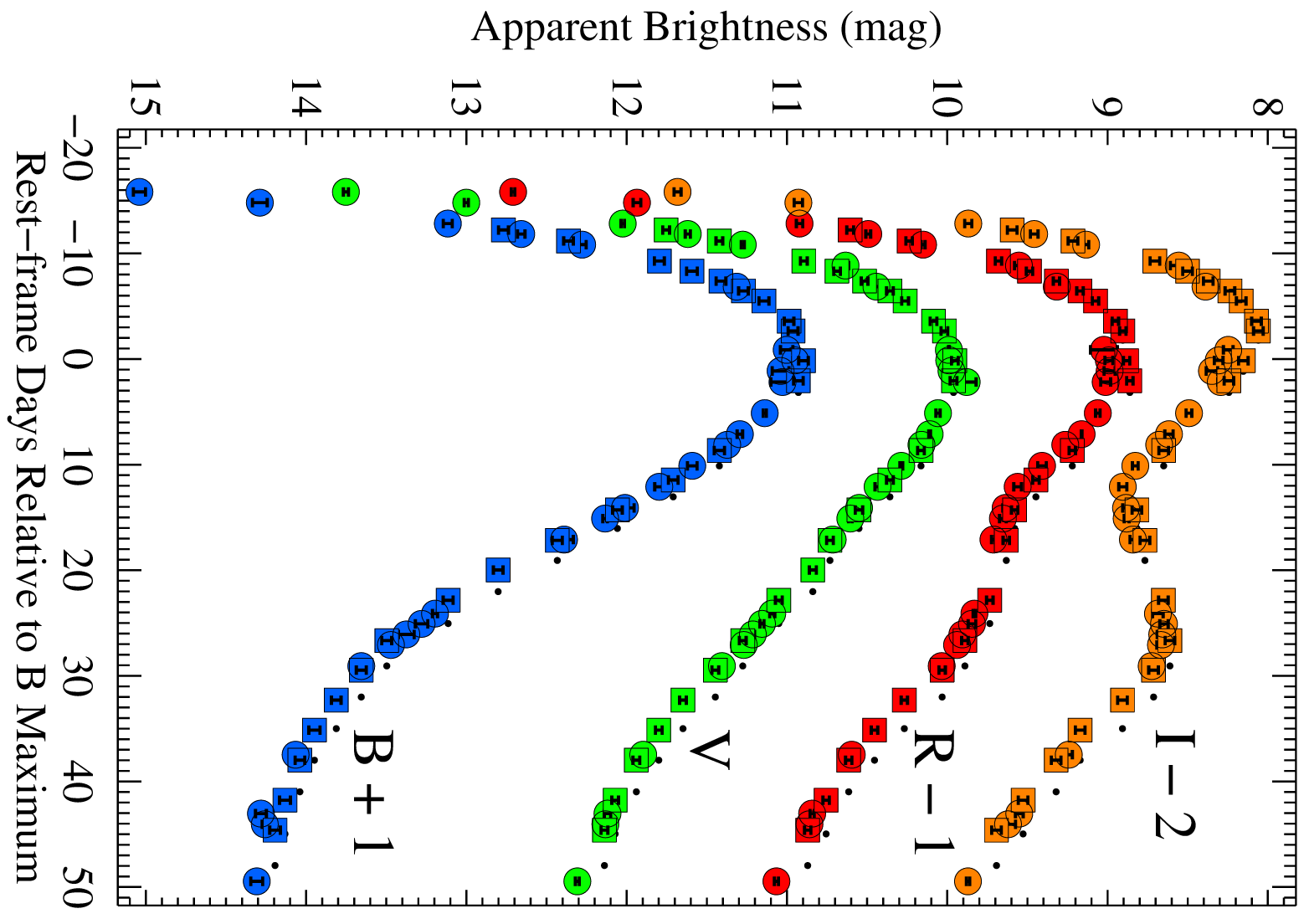}}
\caption{\bvri light curves of SNe~2011by \citep[squares and
  points]{Silverman13} and 2011fe \citep[circles]{Richmond12} with
  offsets noted.  Both SNe have been corrected for Galactic
  extinction.  SN~2011by has been shifted by $-2.91$~mag in all bands,
  to match the apparent peak magnitudes of SN~2011fe, which
  corresponds to a peak absolute magnitude difference of 0.6~mag.  The
  black points are the raw SN~2011by light curves.  The squares are
  the result of shifting the time of maximum $B$ brightness by
  $+1$~day and stretching the light curve by a factor of 0.95,
  consistent with the difference in light-curve shape.  Any
  differences between the light curves can be attributed to inaccurate
  times of maximum, slightly different light-curve shapes, and
  slightly different filter responses
  \citep{Stritzinger02}.}\label{f:lc}
\end{center}
\end{figure}

SNe~2011by and 2011fe have very similar light curves (with $\Delta
m_{15} (B) = 1.14$ and 1.10~mag, respectively, and indistinguishable
rise times; see Figure~\ref{f:lc}).  The SNe also have identical
observed colors.  Using the \citet{Silverman13} data, we find that
SN~2011by peaked at $V = 12.91 \pm 0.01$~mag, while SN~2011fe peaked
at $V = 9.99 \pm 0.01$~mag \citep{Richmond12}.  Correcting for Milky
Way extinction and using our adopted distance moduli, SNe~2011by and
2011fe peaked at $M_{V} = -18.47 \pm 0.36$ and $-19.07 \pm 0.05$~mag,
respectively.  These values are only 1.7$\sigma$ different, but if the
distances are correct, SN~2011by is \about 0.6~mag fainter at peak
than SN~2011fe.  It appears that SN~2011by has a low peak luminosity
for its light-curve shape.

The SNe also have indistinguishable ejecta velocities ($v_{\rm
  Si~II}^{0} = -10$,$300 \pm 200$ and $-10$,$400 \pm 200$~\kms,
respectively).  Figure~\ref{f:by_fe} demonstrates that these two
spectra are virtually identical in the optical.  However, the spectra
diverge at wavelengths $<$2700~\AA, with SN~2011fe having more UV
flux.

\begin{figure}
\begin{center}
\epsscale{1.15}
\rotatebox{0}{
\plotone{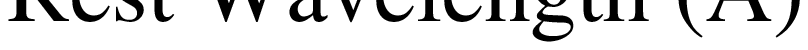}}
\caption{Maximum-light spectra of SNe~2011by (red) dereddened by
  $A_{V} = 0.5$~mag and $R_{V} = 3.1$ and 2011fe
  (black).}\label{f:red}
\end{center}
\end{figure}

Despite the clear similarity in the optical, we investigate if dust
reddening can cause the UV differences.  To make the UV spectra match,
we must assume an extinction of $A_{V} = 0.5$~mag with $R_{V} = 3.1$
and a \citet{Cardelli89} reddening law modified by \citet{Odonnell94}
(Figure~\ref{f:red}).  However, doing this makes the near-UV and
optical continua significantly different.  A non-standard $R_{V}$
cannot account for the difference.  The best explanation is that
SNe~2011by and 2011fe have similar (and probably negligible)
host-galaxy dust reddening, but physical differences that create the
differences in the UV SEDs.

\begin{figure}
\begin{center}
\epsscale{1.1}
\rotatebox{90}{
\plotone{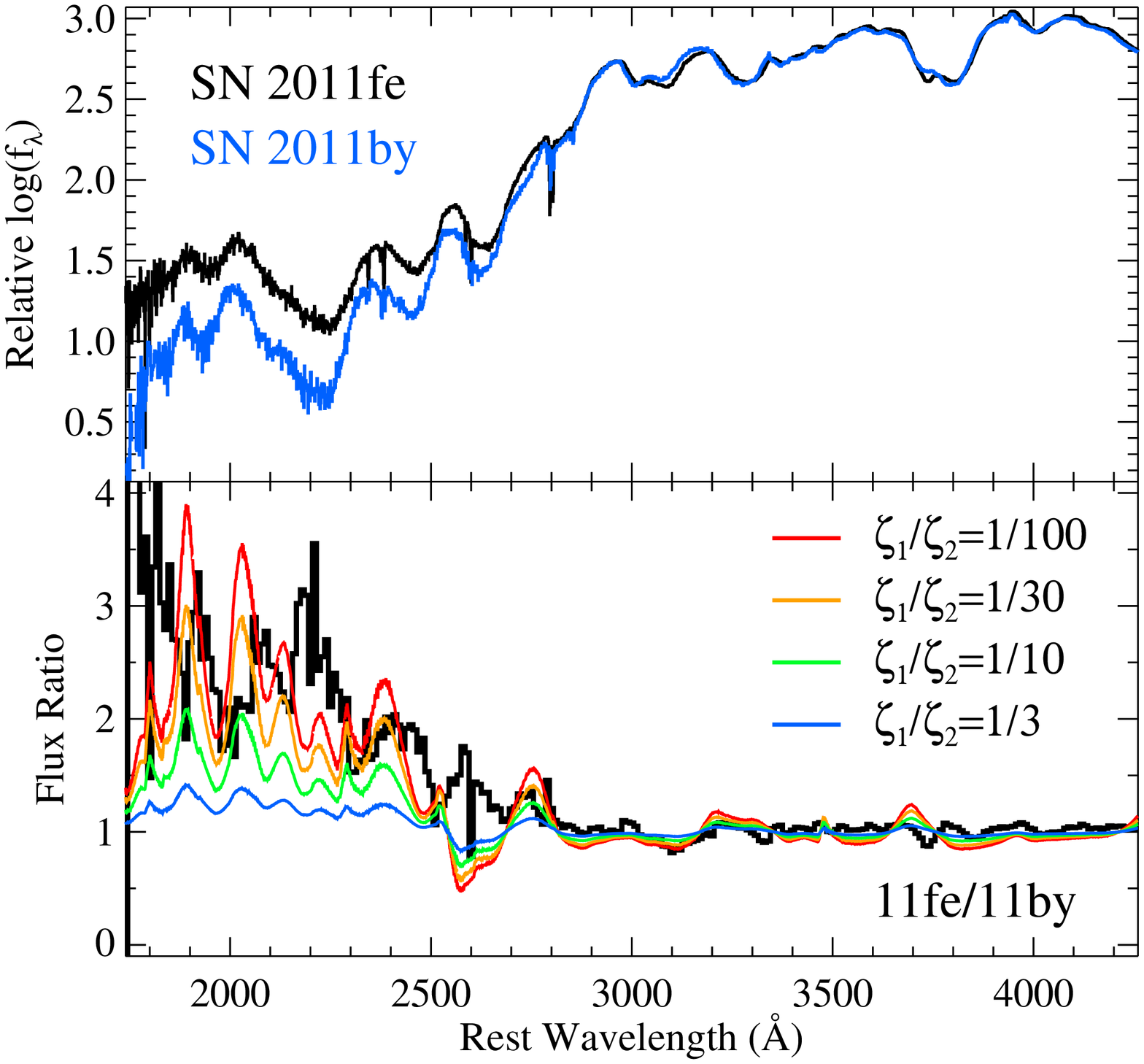}}
\caption{(\textit{Top}) Maximum-light spectra of SNe~2011by (blue) and
  2011fe (black).  (\textit{Bottom}) Flux-ratio spectra for these SNe
  in black.  Overplotted are the average flux-ratio spectra of the
  \citet{Lentz00} $\tau = 15$~day spectra, with differences in the
  metallicity factors, $\zeta$, labeled.}\label{f:spec}
\end{center}
\end{figure}

In Figure~\ref{f:spec}, we compare the UV spectra of the SNe against
each other and examine the flux ratio of the spectra (having scaled
the SNe to match in the optical, and having the same scalings as shown
in Figure~\ref{f:by_fe}).

Since SNe~2011fe and 2011by have very similar light curves and ejecta
velocities, the differences in their far-UV spectra must be
uncorrelated with these parameters and related to another parameter.
Similarly, the cause of the UV difference must not affect the optical
properties of the SN.  For example, mixing of freshly synthesized
Fe-group material to the outer layers should also mix out $^{56}$Ni,
which should change the rise time.  One viable candidate to produce
the UV difference is progenitor metallicity.

\citet{Lentz00} generated several model SN~Ia spectra that cover UV
wavelengths with varying metallicity.  Starting with a single W7
model, they changed the outer-layer metallicity for elements heavier
than O by factors of $\zeta = 1/30$, 1/10, 1/3, 1, and 3.  Increasing
the progenitor metallicity depressed the UV flux.  \citet{Hoflich98}
found the opposite trend, but this was the result of differing density
structures \citep{Lentz00}.  Although the method of \citet{Lentz00} is
far from a complete treatment of how metallicity affects the spectra
of SNe~Ia, the models provide some predictions for short wavelengths.

Although the models generally match the data, they do not precisely
reproduce all spectral features.  By using flux ratios, we avoid some
of the uncertainties in matching the data to the models.  Flux-ratio
spectra focus on the differences between models, which is metallicity
for the \citet{Lentz00} models.  The flux ratios of models with the
same ratio of metallicity factors (i.e., comparing the $\zeta = 1/30$
and 1/3 models and the $\zeta = 1/10$ and 1 models) all have similar
shapes and amplitude.  This means that the flux-ratio spectra, in
principle, can determine differences in metallicity even if the
absolute metallicity is not well determined.  For each ratio of
metallicity factors ($\zeta_{1}/\zeta_{2} = 1/3$, 1/10, 1/30, and
1/100), we produced an average flux-ratio spectrum and present the
spectra in Figure~\ref{f:spec}.  Although the flux ratios remove some
of the uncertainties from the models, more realistic models are
required to produce higher fidelity results.

The model flux-ratio spectra are all approximately 1 (i.e., the
individual spectra are similar) for $\lambda > 2500$~\AA.  Blueward of
2500~\AA, the lower metallicity models have a UV excess relative to
the higher metallicity models.  Moreover, the flux excess increases
with larger differences in metallicity.  For a metallicity factor
ratio of 1/3, the flux difference is relatively small and less than
40\% different at all wavelengths examined.

Some \citet{Lentz00} flux-ratio spectra roughly match the
SN~2011fe/SN~2011by flux-ratio spectrum.  Ascribing the differences in
the UV spectra to differences in progenitor metallicity, the
progenitor of SN~2011by had a higher metallicity than that of
SN~2011fe.  If the \citet{Lentz00} models accurately reproduce the
change in UV continuum for a given difference in metallicity, then the
progenitor of SN~2011by had $>$30 times the abundance for elements
heavier than O than SN~2011fe.  This can be written as
\begin{equation}\label{e:met}
  [{\rm O/Fe}]_{\rm 11by} \lesssim [{\rm O/Fe}]_{\rm 11fe} + 1.5.
\end{equation}

\citet{Timmes03} found that the $^{56}$Ni yield of a SN~Ia should
depend on the metallicity of the progenitor star, deriving a simple
equation relating the progenitor metallicity to the $^{56}$Ni mass,
\begin{equation}\label{e:ni}
  M(^{56}{\rm Ni}) \propto 1 - 0.057 Z / Z_{\sun}.
\end{equation}
For subsolar metallicity, this function is relatively flat.  As an
example, going from $[{\rm O/Fe}] = -1.25$ to 0.25 or $-0.7$ to 0.8
(both $\Delta [{\rm O/Fe}] = 1.5$) would result in $^{56}$Ni mass
differences of 10\% or 50\%, respectively.  Large differences in
$^{56}$Ni mass can only occur if at least one progenitor had above
solar metallicity.

\citet{Arnett82} showed that the peak bolometric luminosity of a SN~Ia
is directly proportional to the amount of $^{56}$Ni generated in the
explosion.  Examining their light curves, SNe~2011by and 2011fe have
consistent rise times.  From their maximum-light spectra, they have
the same optical and near-UV colors.  With the reasonable assumption
that SNe~2011by and 2011fe have the same bolometric corrections, we
have
\begin{equation}
  \frac{M_{\rm 11fe}(^{56}{\rm Ni})}{M_{\rm 11by}(^{56}{\rm Ni})} = 10^{0.4 (M_{V{\rm , 11by}} - M_{V{\rm , 11fe}})},
\end{equation}
and thus $M_{\rm 11fe}(^{56}{\rm Ni}) / M_{\rm 11by}(^{56}{\rm Ni}) =
1.7\err{0.7}{0.5}$.  The uncertainty in the distance to NGC~3972, the
host galaxy of SN~2011by, is the largest uncertainty in the $^{56}$Ni
ratio; better measurements should improve our understanding of these
SNe.  Nonetheless, SN~2011fe generated more $^{56}$Ni than SN~2011by,
consistent with SN~2011fe having lower progenitor metallicity.

If the entire difference is attributed to metallicity, that provides
an estimate for the progenitor metallicity.  Using Equation~\ref{e:ni}
and the 1-$\sigma$ lower bound of the $^{56}$Ni ratio, corresponding
to 1.2, the progenitor of SN~2011by must have been 4$Z_{\sun}$ if the
progenitor of SN~2011fe had solar metallicity.  That is, the
progenitor of SN~2011by almost certainly had a metallicity above
solar.  From Equations~\ref{e:met} and \ref{e:ni}, the best estimate
of the $^{56}$Ni ratio corresponds to metallicities of $[{\rm O/Fe}] =
-0.6\err{0.1}{0.3}$ and 0.9$\err{0.1}{0.3}$ for SNe~2011by and 2011fe,
respectively.  These estimates rely on the \citet{Lentz00} models, and
better models may yield different values.


\section{Discussion \& Conclusions}\label{s:conc}

We presented \textit{HST} maximum-light UV spectra of two SNe~Ia.
SNe~2011by and 2011fe are ``twin'' SNe, having nearly identical light
curves and optical spectra.  SN~2011fe has a higher continuum at
$\lambda < 2500$~\AA\ than SN~2011by and was more luminous.
Considering their remarkable similarity in other respects, the
differences in peak luminosity and UV flux are likely connected.  The
physical cause of both effects must not affect other observables such
as rise time and the optical spectrum.  The only known way to change
both the UV flux and the peak luminosity without affecting other
observables is differing progenitor metallicity.

Comparing the flux-ratio spectrum of SNe~2011by and 2011fe to model
flux-ratio spectra, we determined that SN~2011by had a higher
metallicity with $\Delta [{\rm O/Fe}] \gtrsim 1.5$.  The SNe had peak
absolute magnitudes that differed by \about 0.6~mag, which corresponds
to a ratio in their $^{56}$Ni masses of $1.7\err{0.7}{0.5}$; the
largest contribution to the uncertainty is from the distance to
SN~2011by.  Given the relative metallicity difference and $^{56}$Ni
mass ratio, we estimate that the progenitors of SNe~2011by and 2011fe
had supersolar and subsolar metallicities, respectively.  Improved
modeling that includes more realistic physics (e.g., multi-dimensional
models, improved flame physics) will be necessary to accurately and
precisely measure progenitor metallicities.

SNe~2011by and 2011fe were hosted by Sbc and Scd galaxies which have
gas-phase metallicities of $12 + \log({\rm O/H}) = 8.97$ and 9.12 from
SDSS spectra \citep[using a solar value of 8.86]{Prieto08:met},
respectively.  However, \citet{Bresolin07}, performing a more detailed
analysis of M101, found a metallicity of 8.93 (using a solar value of
8.66).  Using the \citet{Bresolin07} metallicity and metallicity
gradient, \citet{Stoll11} determined that M101 had a metallicity of
8.45 at the position of SN~2011fe.  A similar measurement must be made
for SN~2011by, but taken at face value, the progenitor sites of the
two SNe could differ in metallicity by \about 0.5 and in the direction
we found from the SNe.  The metallicity at the radius of the SN may
roughly indicate the progenitor metallicity.

There are three other reported measurements for SN~Ia progenitor
metallicity, but there is no differential metallicity measurement
similar to what we have presented here.  On the face of it,
\citet{Taubenberger08} found that the progenitor of SN~2005bl, a
low-luminosity SN~Ia, had subsolar metallicity based on the strength
of optical Fe lines.  By measuring the relative flux of X-ray lines
from their SN remnants, \citet{Badenes08:met} and \citet{Park13}
determined that the progenitors of Tycho's and Kepler's SNe both had
$Z \approx 3 Z_{\sun}$ with relatively large uncertainties.  It
appears that the progenitors of SNe~2005bl and 2011fe had similar
(subsolar) metallicities, while the progenitors of SN~2011by, Tycho's
SN, and Kepler's SN had similar (supersolar) metallicities.

Using the SNLS sample of SNe~Ia and proxies for $^{56}$Ni mass (peak
SN luminosity in multiple bands) and metallicity (host-galaxy
luminosity), \citet{Howell09} also found a trend between the two
quantities consistent with the predictions of \citet{Timmes03}.
Similarly, several authors have found that SNe~Ia in more massive
galaxies have higher luminosity (after light-curve shape corrections)
than those in less massive galaxies \citep{Kelly10, Lampeitl10:host,
  Sullivan10}.  These results are consistent with those found here,
where the SN with a higher metallicity progenitor, SN~2011by, was
fainter than the SN with the same light-curve shape but lower
metallicity progenitor, SN~2011fe.

The average metallicity of SN~Ia progenitors must increase with cosmic
time and decreasing redshift.  This, in turn, should produce fainter
SNe for the same light-curve shape.  Not accounting for this effect
could bias SN~Ia cosmological measurements.  Specifically, SNe~Ia at
high redshift should have, on average, lower metallicity and higher
luminosity than local comparison SNe, resulting in an underestimate of
the effect of acceleration.  Using host-galaxy mass as a proxy for
progenitor metallicity may remove this potential systematic bias.  The
effect of progenitor metallicity is relatively small, but accurately
constraining the properties of dark energy demands attention to small
systematic effects.

\begin{acknowledgments} 

  {\it Facilities:} \facility{HST(STIS)}

  \bigskip Supernova research at Harvard is supported by NSF grant
  AST--1211196.  We thank the KITP for their hospitality, where this
  research was supported in part by NSF grant PHY11--25915.

  We thank W.\ Hillebrandt, M.\ Kromer, T.\ Piro, J.\ Silverman, M.\
  Stritzinger, and S.\ Taubenberger for useful discussions, the
  anonymous referee for helpful comments, and J.\ Silverman for
  providing data.

  This research has made use of the NASA/IPAC Extragalactic Database
  (NED), which is operated by the Jet Propulsion Laboratory,
  California Institute of Technology, under contract with NASA.
  Support for GO Program number 12592 was provided by NASA through a
  grant from the Space Telescope Science Institute, which is operated
  by the Association of Universities for Research in Astronomy,
  Incorporated, under NASA contract NAS5--26555.

\end{acknowledgments}

\bibliographystyle{../fapj}
\bibliography{../astro_refs}


\end{document}